\documentclass[letterpaper,11pt]{article}

\usepackage{jheppub} 

\usepackage{graphicx}
\usepackage{epstopdf}
\usepackage{mathtools}
\usepackage{amsmath, amsthm, amssymb}

\usepackage[usenames,dvipsnames,svgnames,table]{xcolor} 
\usepackage{tikz}
\usepackage{verbatim}
\usepackage{bm}
\usepackage{bbold}
\usepackage{caption}
\usepackage{subcaption}

\usepackage[utf8]{inputenc}
\usepackage[T1]{fontenc}

\usepackage[smalltableaux]{ytableau}

\newcommand{\be}{\begin{eqnarray}}
\newcommand{\ee}{\end{eqnarray}}
\newcommand{\bea}{\begin{eqnarray}}
\newcommand{\eea}{\end{eqnarray}}

\newcommand{\bn}{\begin{enumerate}}
\newcommand{\en}{\end{enumerate}}

\def\CN{{\cal N}}

\def\a{\alpha}
\def\b{\beta}

\def\half{\frac{1}{2}}

\def\vev#1{\langle #1 \rangle}

\def\Tr{{\rm Tr}}

\def\vec#1{\bm{#1}}

\title{Notes on S-folds and $\mathcal{N}=3$ theories.}

\author[a]{Prarit Agarwal,}
\author[b]{Antonio Amariti,}

\affiliation[a]{Department of Physics,Seoul National University \\Seoul, South Korea}
\affiliation[b]{ Physics Department, The City College of the CUNY,\\
160 Convent Avenue, New York, NY 10031, USA}

\emailAdd{agarwalprarit@gmail.com}
\emailAdd{aamariti@ccny.cuny.edu}

\abstract{We consider D3 branes in presence of an S-fold plane.
The latter is a non-perturbative object, arising from the combined projection of 
an S-duality twist and a discrete orbifold of the R-symmetry group.
This construction naively gives rise to 4d $\mathcal{N}=3$ SCFTs.
Nevertheless it has been observed that in some cases supersymmetry is enhanced
to $\mathcal{N}=4$.
In this paper we study the explicit counting of degrees of freedom arising from vector multiplets associated to strings suspended between the D3 branes probing the S-fold. We propose that, for trivial discrete torsion, there is no vector multiplet associated to $(1,0)$ strings stretched between a brane and its image.  We then focus on the case of rank 2 $\CN=3$ theory that  enhances to $SU(3)$ $\CN=4$ SYM, explicitly spelling out the isomorphism between the BPS-spectrum of the manifestly $\CN=3$ theory and that of 
three D3 branes in flat spacetime. 
Subsequently, we consider 3-pronged strings in these setups and show how wall-crossing in the S-fold background implies wall crossing in the flat geometry. This can be considered a consistency check of the \emph{conjectured} SUSY enhancement. We also find that the above isomorphism implies that a $(1,0)$ string, suspended between a brane and its image in the S-fold, corresponds to a 3-string junction in the flat geometry. This is in agreement with our claim on the absence of a vector multiplet associated to such $(1,0)$ strings.  This is because the 3-string junction in flat geometry gives rise to a $1/4$-th BPS multiplet of the $\CN=4$ algebra. Such multiplets always include particles with spin $>1$ as opposed to a vector multiplet which is restricted by the requirement that the spins must be $\leq 1$.    
}
\begin{document} 
\maketitle
\flushbottom

\section{Introduction}

In a recent paper \cite{Aharony:2015oyb} the properties of 4d SCFTs with twelve supercharges, i.e. $\mathcal{N}=3$ theories, have been discussed starting from the constraints imposed by the superconformal algebra. 
Surprisingly an explicit realization of such theories has been obtained in 
\cite{Garcia-Etxebarria:2015wns},
starting from F-theory
\footnote{See also \cite{Ferrara:1998zt} for a construction based on supergravity.}.

This discovery attracted quite much interest \cite{Beck:2016lwk,Cordova:2016xhm,Nishinaka:2016hbw,Argyres:2016xua,Aharony:2016kai,Morrison:2016nrt,Imamura:2016udl}. One of the main reason is that $\mathcal{N}=3$ 
have been overlooked in the past because they are necessarily non-lagrangian

The F-theory  construction of \cite{Garcia-Etxebarria:2015wns} is based on the combined action of an S-duality and an R-symmetry twist.
This action corresponds in string theory to introducing a non-perturbative object, generalizing the 
action of an orientifold plane. This object has been named S-fold in \cite{Hull:2004in}.
It carries  charges under the R and the NS sectors in string theory
and, as discussed in \cite{Garcia-Etxebarria:2015wns}, there is a classification in terms of 
discrete choices of the torsion classes.

The discrete torsion carried by the S-folds has been further discussed in
\cite{Aharony:2016kai}, where it was also shown that in the rank 2 case and in absence of discrete torsion 
there is a non-perturbative enhancement of supersymmetry from the manifest $\mathcal{N}=3$ to  $\mathcal{N}=4$.
Three cases have been observed, corresponding to $SU(3)$, $SO(5)$ and $G_2$ gauge groups respectively.
The realization of such gauge groups is in an unconventional basis
because the roots carry 
both electric and magnetic charges
under the manifest $U(1)^2$ subgroup of the gauge symmetry

In this paper we further elaborate on the structure of this enhancement for the $SU(3)$
case.  We develop a dictionary by mapping the masses and the charges associated to the $(p,q)$ strings connecting the D3 branes in the S-fold geometry and the corresponding 
states in $SU(3)$ $\mathcal{N}=4$ theory.
We use this mapping to compare more complicated states, arising out of 
3-pronged junctions. We study the masses of such states in the S-fold geometry and discuss the relation with the manifest central charge
\footnote{Here we refer to the central charge of the $\mathcal{N}=3$ theory as the \emph{manifest} central charge because the theory is actually $\mathcal{N}=4$ and
in the S-fold geometry the second central charge is \emph{hidden}.} 
appearing in this construction.
We also study the wall crossing in the S-fold geometry and compare with the $\mathcal{N}=4$
case.

The paper is organized as follows.
In section \ref{sec:summary} we give a review of the recent developments in the study of $\mathcal{N}=3$ theories.
In section \ref{sec:Countingdof} we discuss the charges of $(p,q)$ 
strings in the S-fold geometry.
In section \ref{sec:BPSpq} we compute the central charges and the masses of such strings, developing the dictionary between the S-fold case and the flat $\mathcal{N}=4$ geometry.
In section \ref{3-pronged} we introduce the three pronged junctions. We compute the 
associated walls of marginal stability in the rank 2 S-fold case and 
discuss relation with the $SU(3)$ $\mathcal{N}=4$ theory.
In section \ref{sec:further} we conclude by discussing open problems and further directions.
\\
\\
{\bf Note Added}: While this paper was under preparation, \cite{Imamura:2016udl} appeared on arXiv. This paper studies junctions in rank 2 S-folds leading to the enhancement of SUSY. Despite the apparent 
similarity of the two approaches our analysis is actually complementary. 
While \cite{Imamura:2016udl} studies junctions in the flat case to infer properties of BPS states in the 
S-fold geometry, here we study junctions in the S-fold geometry and map such states 
to the flat case.

\section{A brief summary of developments so far}
\label{sec:summary}

Inspired by recent developments in the understanding of non-lagrangian systems, the authors of \cite{Aharony:2015oyb} investigated the possibility of 4d SCFTs with twelve supercharges. These theories have to be necessarily non-lagrangian. This is because rigid 4d $\CN=3$ SUSY has a unique non-trivial representation when the spin of its components is restricted to be $\leq 1$.  This representation is the vector multiplet, which consists of a single spin-1 gauge field, four Weyl spinors and 6 real scalars. Therefore, any $\CN=3$ lagrangian has to be written in terms of these $\CN=3$ vector multiplets.  However, the $\CN=3$ vector multiplet is identical to the $\CN=4$ vector multiplet and therefore it is invariant under 16 supersymmetries. Thus a lagrangian description of $\CN=3$ will also be invariant under $\CN=4$ supersymmetry. Genuinely, $\CN=3$ systems therefore are not expected to be described by a lagrangian.

Upon considering the constraints put on $\CN=3$ theories by $\CN=2 \subset \CN=3$ SUSY and by comparing them to the properties of $\CN=4$ SYM ( when thought of as trivial examples of $\CN=4$ SYM), the authors of \cite{Aharony:2015oyb} found that the central charges $a$ and $c$ of these theories must be equal to each other.  
They also studied the space of marginal deformations of these systems and concluded that these theories can not have any $\CN=3$ or $\CN=2$ preserving exactly marginal deformations.  To do so they looked for a SUSY preserving, R-symmetry invariant, scalar operator of dimension 4 in the spectrum of operators in $\CN=3$ SCFTs. SUSY invariance of this operator implies that it must be the top component of a SUSY multiplet. They were able to show that there are only two such SUSY multiplets,  both containing an extra conserved supercurrent. Any such operators if present in the theory will therefore imply that there is an extra supercurrent . Such a theory will therefore have $\CN=4$ SUSY.  Hence genuinely $\CN=3$ preserving marginal deformations are not possible. Similar analysis can also be found in \cite{Cordova:2016xhm}. A marginal deformation that preserves only $\CN=1$ SUSY might be possible, although such a possibility was not explored in \cite{Aharony:2015oyb, Cordova:2016xhm}.

It was also shown in \cite{Aharony:2015oyb}, that $\CN=3$ theories do not admit any non-R global symmetries. This is because the $\CN=3$ supermultiplets containing conserved global non-R currents also contain an extra conserved supercurrent, in fact they are the same multiplets that contain the  marginal deformations mentioned in the previous paragraph. Basically, the extra bosonic currents sitting in these multiplets combine with the $SU(3)_R \times U(1)_R$ symmetry to give the $SU(4)_R$ symmetry of $\CN=4$ algebra, while the fermionic currents combine with the $\CN=3$ supercurrents to give the multiplet of $\CN=4$ supercurrents.

Around the same time, an F-theory construction of $\CN=3$ theories was proposed in \cite{Garcia-Etxebarria:2015wns}. Their construction is based on a $k$-fold (called the S-fold in \cite{Aharony:2016kai}) generalization of the F-theory lift of the type-IIB orientifolds. Recall, that  4d compactification of F-theory can be defined in terms of M-theory compactified on an elliptically fibered $CY_4$ and taking a limit in which the fiber shrinks to size zero. $\CN=4$ theories with orthogonal and symplectic gauge groups can be realized through M-theory compactification on $\mathbb{R}^{1,2} \times (\mathbb{C}^3 \times T^2)/\mathbb{Z}_2$. The F-theory lift of the orientifold is such that the $\mathbb{Z}_2$ action on the elliptic fiber in $CY_4$ is realized through an appropriate embedding in its $SL(2,\mathbb{Z})$ transformation. 
This picture corresponds to M-theory on $\mathbb{R}^{1,2} \times (\mathbb{C}^3 \times T^2)/\mathbb{Z}_k$.  
Now, the torus, $T^2$, admits $Z_k$ symmetry only when $k=1,2,3,4,6$ \cite{Dabholkar:2003xi, Nilse:2006jv}. Thus the above orbifold is only possible for these values of $k$. 
Moreover, for $k=3,4,6$, the complex structure of $T^2$ cannot be arbitrary but must be set equal to $\tau=e^{2i\pi/k}$.  The finite values of $\tau$ indicate that this is a non-perturbative background of the 10d string theory. The complex structure of the torus generically becomes a coupling of the 4d theory. The fact that for $k=1,2$ the complex structure of the torus can be arbitrary then reflects the fact that the coupling constant for $\CN=4$ theories is a marginal parameter and can be tuned to any desired value. It was shown in \cite{Garcia-Etxebarria:2015wns} that for $k=3,4,6$ the 4d theory has only $\CN=3$ SUSY. The requirement of fixed value for the complex structure of the torus then fits well with the absence of any $\CN=3$ preserving marginal operators in these theories.  
 In the F-theory limit, this set-up gives rise to a stack of $n$ D3 branes probing a six dimensional transverse space $\mathbb{C}^3/\mathbb{Z}_k$. The  moduli space of the low energy 4d theory must then be $(\mathbb{C}^3/\mathbb{Z}_k)^n/S_n$, where $S_n$ is the permutation group of $n$ objects. 

In \cite{Garcia-Etxebarria:2015wns}, it was pointed out that along with $k$, there are additional parameters that label the S-folds described above. These labels correspond to discrete torsion \cite{Witten:1998xy}. In \cite{Aharony:2016kai}, the theory obtained by considering $n$ D3 branes probing an $S$-fold with $\mathbb{Z}_k$ twist, was found to be related to the complex reflection groups $G(k,p,n)$, with $p$ being a divisor of $k$.  These are generalization of Euclidean reflections to  a complex vector space with hermitian inner product \cite{Shephard01011952,shephard1953unitary,shephard1954finite}. For our purposes here, they can be thought of as generalizations of the Weyl group of Lie algebras. Indeed the complex reflection group $G(1,1,n)$ is isomorphic to $S_n$ which is the Weyl group of the $A_{n-1}$ Lie algebra. This is consistent with the fact that when $k=1$, the S-fold is trivial. We therefore get a stack of $n$ D3 branes on a flat background and the low energy theory is given by an $\CN=4$ $SU(n)$ gauge theory. Similarly, $G(2,1,n)$ is isomorphic to the Weyl group of $B_n$ and $C_{n}$ type Lie algebras. In this case, the $S$-fold becomes an orientifold with an appropriate discrete torsion switched on. On the other hand, $G(2,2,n)$ is the Weyl group of $D_n$ Lie algebras. This time the S-fold becomes an orientifold but without any discrete torsion. As was argued in \cite{Aharony:2016kai}, the label $p$ in $G(k,p,n)$ can be associated to discrete torsion even when $k=3,4$ and $6$, such that there is no discrete torsion when $p=k$. It was also argued that for  $p>1$, there will also be a $\mathbb{Z}_p$ discrete gauge symmetry acting on the theory.  Keeping all the labels explicit, we will call such an object as $S_{k,l}$-fold, where $l=k/p$.

The ring of Coulomb branch operators of these theories (with ungauged $\mathbb{Z}_p$ symmetry) will then be given by the ring of invariants of $G(k,p,n)$ . By considering the relation between the central charges and the  dimension of Coulomb branch operators, it was shown in \cite{Aharony:2016kai} that the central charges $a$ and $c$ of these theories are given by
\be
4a=4c=2k\sum_{m=1}^{n-1} m + 2nl-n = k n^2 + n(2l-k-1) , \ l=\frac{k}{p} \ .
\label{eq:N=3centralcharges}
\ee
Now, we recall that for non-lagrangian $\CN=2$  theories we can define an effective number of vector and hypermultiplets given by 
\be
c = \frac{2n_v + n_h}{12} \quad \text{and} \quad a = \frac{5n_v + n_h}{24} \ .
\ee
We can use this to define an effective number of $\CN=2$ vector and hypermultiplets in the present theories. We then find that 
\be
n_v = n_h = 4a=4c =  k n^2 + n(2l-k-1) \ .
\label{eq:effVec}
\ee
The fact that $n_v = n_h$ in our theories should not be surprising. This is because the $\CN=3$ analog of such quantities will necessarily be given by the effective number of $\CN=3$ vector multiplets ($\tilde{n}_v$) . Since each such $\CN=3$ vector multiplet will split into an $\CN=2$ vector multiplet and a hypermultiplet, they will therefore appear in pairs thereby giving $n_v=n_h = \tilde{n}_v = 4a =4c$. When $k=p=2$, we find that $\tilde{n}_v = 2n^2-n$ which is the number of vector multiplets we require to form an adjoint representation of the $SO(2n)$ gauge symmetry in the $\CN=4$ theory obtained from this orientifold. Similarly, when $k=2,\ p=1$, $\tilde{n}_v = 2n^2+n$ which is equal to the number of generators of the $Sp(n)$ (or equivalently $SO(2n+1)$) gauge symmetry described by such orientifolds.  It is not clear if a similar interpretation of $\tilde{n}_v$ in terms of a root system is possible when $k \neq 1,2$.
 
 As we will try to demonstrate later, one possible way to understand $\tilde{n}_v$ is in terms of vector multiplets arising from $(1,0)$ strings stretched between the various branes. However, because of the non-trivial embedding of $\mathbb{Z}_k$ inside $SL(2,\mathbb{Z})$, not all of these vector multiplets will have mutually local electromagnetic charges with respect to the $U(1)^n$ symmetries generated by branes in the stack. This lack of mutual locality then makes it impossible to be able to write a lagrangian that simultaneously describes all of them as elementary degrees of freedom (d.o.f.) of the theory.

\section{An explicit counting of the degrees of freedom}
\label{sec:Countingdof}
Here we discuss the 
BPS states of the rank 2 $\mathbb{Z}_3$ S-fold.
We start our analysis from the well known case of $n$ D3 branes probing an
O3 plane. This corresponds to the $\mathbb{Z}_2$ projection in the language of the S-fold.
This picture is useful to develop the intuition on the strings stretched between the branes 
in the orbifolded geometry. 
Then we generalize the description to the $\mathbb{Z}_k$ case.
We construct the charge lattice of $(p,q)$ strings connecting the D3 branes
(and their images)  probing the S-fold.

\subsection{A lesson from the orientifold: the $Z_2$ projection}

\begin{figure}[h]
	\centering
	\includegraphics{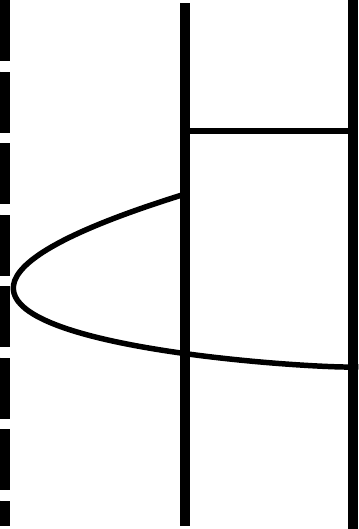}
	\caption{Strings stretched between two D3 branes on top of an $O3$-plane}
	\label{fig:Orientifold}	
\end{figure}

We consider a stack of $n$ D3 branes on top of the $O3^-$ plane. This gives rise to an $\CN=4$, $O(n)$ gauge theory. The $\mathbb{Z}_2$ orbifolded geometry of the space transverse to the D3 brane implies we can construct two non-homotopic paths between any pair of D3 branes. One way to think about this is that the if one path takes us from the first D3 brane to the second D3 brane, then the second path takes us from the first D3 brane to the image of the second D3 brane (see figure \ref{fig:Orientifold} for a cartoon of strings 
connecting two D3 branes on top of an orientifold). 
The spectrum of BPS-states in the theory is obtained by $(p,q)$strings stretched between the branes along the homotopically distinct paths that connect them. Each such state can be assigned a 4-vector of electromagnetic charges with respect to the $U(1)^2$ gauge symmetry generated by the system of two branes. We will label these as $(e_1,m_1;e_2,m_2)$. The $(p,q)$string along the trivial path gives rise to a state with charges $(p,q;-p,-q)$ while the string stretched from the first brane to the image of the second brane will correspond to a BPS-state with charges $(p,q;p,q)$.

 We now consider the vector multiplet states associated to F1 strings stretched between the D3 branes. This gives an electromagnetic charge $(e_i,m_i;e_j,m_j)=(1,0;-1,0)$ for the state coming from the string connecting the $i$-th D3 brane to the $j$-th D3 brane and a charge 
 $(e_i,m_i;e_j,m_j)=(1,0;1,0)$ for the state coming from the string connecting the $i$-th D3 brane to the image of $j$-th D3 brane. In order to complete the set of states in the adjoint representation, we'll also have to consider the d.o.f given by F1 strings going from a D3 brane to itself. Depending upon the type of the orientifold ($O3^+$ vs $O3^-$ and $\widetilde{O3}-$), we might or might not have the vector  multiplets corresponding to a $\pm(1,0)$ string going from a D3 brane to its image. In the case of $O3^+$-planes, such vector multiplets are there and we get an $Sp(n)$ gauge group when all the branes are made coincident with the orientifold.  
\subsection{Generalization to the $\mathbb{Z}_k$ case}
We now generalize the above description to the case of an $S_{k,l}$-fold.
This time we consider a stack of $n$ $M2$-branes on $(\mathbb{C}^3 \times T^2)/\mathbb{Z}_k$. In the F-theory limit, this gives us $n$ D3 branes probing $\mathbb{C}^3/\mathbb{Z}_k$. Orbifolding will give rise to a k-fold connected geometry or equivalently $k-1$-images for each brane. Thus a pair of D3 branes will be connected by $k$ homotopically distinct paths. $(p,q)$ strings stretched between the branes along these paths will then give rise to the BPS spectrum of our theory. Once again, let us consider the vector multiplets associated to $(1,0)$ strings stretched between a pair of branes branes.  This gives us rise to $2k$ elementary real d.of. 
We will also have to consider the states arising from strings going from a D3 brane to itself or to one of its images. Depending upon the value of $l$, some of these states might get projected out. We \emph{conjecture} that when $l=1$, all the states going from a D3 brane to one of its images will be projected out and only the states going from a D3 brane to itself contribute. On the other hand, when $l=k$, none of these states get projected out and all of them will have to be taken into account . The state going from a D3 brane to it-self give real d.o.f , while the state going from a D3 brane to one of its image gives a complex d.o.f. Therefore, when $l=1,k$, we get $2l-1$ real d.o.f living on a brane. As shown in \cite{Aharony:2016kai}, $l=1$ and $l=k$ are the only values for which an $\CN=3$ S-fold exists, we therefore do not worry about the rule for projecting out states when $1 < l <k$. The total number of d.o.f in the theory then becomes
\be
   \text{ \# of d.o.f}  = 2k \frac{n(n-1)}{2} + (2l-1)n  = k n^2 + (2l-k-1)n  , \ l=1,k
\ee  
which agrees with \eqref{eq:N=3centralcharges}, \eqref{eq:effVec}.

\section{$1/2$ BPS states and SUSY enhancement}
\label{sec:BPSpq}

It  was \emph{conjectured} in  \cite{Aharony:2016kai} that 4d rank 2 $\CN=3$ theories with $l=1$ and $k=3,4,6$ are actually invariant under one more supercharge which makes them isomorphic to 4d $\CN=4$ SYM with gauge group $G$ being $SU(3)$, $SO(5)$ and $G_2$ respectively. One hint that this might be the case is already provided by the fact that the corresponding complex reflection group $G(k,k,2)$ for $k=3,4$ and $6$ is isomorphic to the Weyl group of $SU(3),\ SO(5)$ and $G_2$ respectively. 
More convincing evidence for this enhancement was provided in  \cite{Aharony:2016kai}.

In this section we study the matching of $(p,q)$ strings associated to $1/2$-BPS states of 
$\CN=4$ SYM to those obtained from rank 2 $\CN=3$ theories exhibiting this SUSY enhancement. We restrict our analysis to the case of the $S_{3,1}$-fold.
We start this section by fixing some notations and the glossary of relevant formulae
for the charge and the mass of a $(p,q)$ string.
Then we study the spectrum of such dyonic states in presence of an $S_{3,1}$-fold
and obtain the explicit mapping of this spectrum 
with that of the corresponding $SU(3)$ $\CN=4$ theories.

\subsection{Charges and masses of $(p,q)$ strings}

Here we discuss some relevant formulae that will be useful in the rest of the discussion.
More specifically we review the formulae for the tension
of a $(p,q)$string, its electric and its magnetic charge in terms of the roots and the coroots,
the central charge and the BPS mass formula.
The tension of a $(p,q)$string is given by 
\be	
T_{p,q} = \frac{\sqrt{g_s}}{2 \pi \alpha'} |p + q \tau | \ ,
\label{eq:stringtension}
\ee
where 
\be
\tau = \frac{i}{g_s} + \frac{\chi}{2 \pi} = \frac{4 \pi i}{g_{YM}^2} + \frac{\theta_{YM}}{2 \pi} \ .
\label{eq:StringCoupling}
\ee
Let us consider a rank r theory and let the electric and magnetic quantum numbers of a given state in this theory be $(n_e^a, n_m^a)$ for $a = 1,\hdots,r$. Also, let $\vev{\phi^I}= v_a^I H^a$ for $I= 1,\hdots, 6$ be the vev of the $I$-th real scalar in the $\CN=3, 4$ vector multiplet. Here $H^a$ is the $a$-th Cartan generator of the gauge group. We will use $\beta^a$ to denote the $a$-th simple root and $\beta^{*}_a$ will denote the corresponding coroot. Then the particle created by this state has the following electric and magnetic charges
\be
\begin{split}
Q_E^I &= g_{YM} \sum_{a=1}^{r} (n_e^a \beta^a v^I_a +\frac{\theta_{YM}}{2\pi} n_m^a \beta^{*}_a v^I_a) \ , \\
Q_M^I &= \frac{4 \pi}{g_{YM}} \sum_{a=1}^{r} n_m^a \beta^{*}_a v^I_a \ .
\end{split}
\ee 
Let $\vec{Q_E}$ and $\vec{Q_M}$ be six dimensional (real) vectors with their $I$-th components being $Q_E^I$ and $Q_M^I$, respectively, as given above. 
We can now use R-symmetry to rotate the vevs $v^I$ such that they are non-zero only when $I=1,2$. In the brane picture this corresponds to considering all the branes as coplanar.  The central charge is then given by
\be
\begin{split}
Z = \vec{Q_E} + i \vec{Q_M} =g_{YM}\sum_{a=1}^r (n_e^a \beta^a z_a + \tau n_m^a \beta^{*}_a z_a) \ ,
\end{split}
\ee 
with $z_a =  (v_a^1+ i v^2_a) $. The mass of the corresponding BPS state becomes 
\be
M_{BPS}^2 = |Z|^2 = |\vec{Q_E}|^2 + |\vec{Q_M}|^2 + 2 |\vec{Q_E} \times \vec{Q_M}| \ .
\ee
Corresponding to $z_a$, the position, $z_a'$, of the a-th brane which is given by $z_a'= 4 \pi^{\frac{3}{2}} \alpha' z_a$. 
If a $(p,q)$string ends on the $n$-th image of a brane, ($0 \leq n < k$ for an $S_{k,l}$-fold), then the electromagnetic quantum numbers $(n_e, n_m)$ that it imparts to the BPS state are given by $n_e + n_m \gamma = (p+q \gamma) \gamma^n$ \footnote{Recall that $\gamma: \gamma^k=1$} .

\subsection{$1/2$-BPS states of rank 2 S$_{3,1}$-fold and
enhancement to $SU(3)$ $\CN=4$ SYM}

 The $U(1)^2$ subgroup of the gauge symmetry, manifest in the $\CN=3$ construction, should not be confused with the Cartan subgroup of $G$. In fact, the manifest $U(1)^2$ symmetry is realized through some ``unconventional'' mixing between the electric and magnetic $U(1)$'s of the theory. This implies that the roots of the gauge group carry both electric and magnetic charges with respect to $U(1)^2$ that is visible in the $\CN=3$ formulation. For example, in this basis, the electromagnetic charges of the 6 non-zero roots of $SU(3)$ theory were given by
\be
\begin{split}
(e_1,m_1;e_2,m_2) = \pm(1,0;-1,0) , \ \pm(0,1;1,1), \ \pm(1,1;0,1) \ ,
\end{split} 
\label{eq:SU(3)N=3roots}
\ee

The spectrum of dyonic states of the $\CN=3$ theories above, matches with that of the corresponding $\CN=4$ theories  , when the gauge coupling of the $\CN=4$ theories is given by $\tau_{YM} = -\frac{1}{1+\gamma}$ with $\gamma^k=1$ \cite{Aharony:2016kai}. In the rest of this section we make this map explicit for the case of $k=3$ and $l=1$. This case is expected to be isomorphic to $\CN=4$ $SU(3)$ SYM. 

We establish the above mentioned isomorphism by thinking of the $\CN=4$ $SU(3)$ SYM as being the low energy theory of a stack of three parallel D3 branes. There is a $\mathbb{C}^3$ transverse to these branes and we can use the $SO(6)_R$ of the theory to localize the branes on a complex plane inside $\mathbb{C}^3$. The generic coordinates of the three branes on this complex plane are then given by $0, \tilde{z}_1$ and $\tilde{z}_2$ respectively.  The roots are then generated by $(1,0)$ strings stretched between the branes. Correspondingly, their central charges will be proportional to $\pm \tilde{z}_1$, $\pm\tilde{z}_2$ and $\pm(\tilde{z}_1-\tilde{z}_2)$. We can then choose a system of positive roots given by $\tilde{z}_1 - \tilde{z}_2$, $\tilde{z}_2$ and $\tilde{z}_1$ with the simple roots being $\tilde{z}_1 - \tilde{z}_2$ and $\tilde{z}_2$. The same exercise in the $\CN=3$ theory with the roots as given in \eqref{eq:SU(3)N=3roots}, tells us that the charge vector $(1,0;-1,0)$, $(0,1;1,1)$ and $(1,1;0,1)$ can be chosen to be the positive roots with  $(1,0;-1,0)$ and $(0,1;1,1)$ being the simple roots. The isomorphism between the two formulations will preserve the central charge of the various BPS states. This is because the mass of a BPS state is given by the absolute value of its central charge. Mass being a physical observable, should not change irrespective of whether we use the $\CN=3$ construction or the more familiar construction in terms of a stack of D3 branes. We will further assume that not only the absolute value but also the phases of the central charges match. This implies that a 
possible isomorphic map between the states of the two theories can be realized by mapping $(1,0;-1,0)$ to $\tilde{z}_1 - \tilde{z}_2$ and $(0,1;1,1)$ to $\tilde{z}_2$. This tells us that, 
\be
\begin{split}
\tilde{z}_1 - \tilde{z}_2 & \propto z_1-z_2 \ , \\
              \tilde{z}_2 & \propto \omega z_1 + (1+\omega) z_2  \ , \\
              \tilde{z}_1 &\propto -\omega^2 z_1 + \omega z_2 \ , 
\end{split}
\label{eq:SU(3)rootmatching1}
\ee 
here $\omega$ is the cube root of unity such that $\omega^3=1$. Let us, for a moment, be agnostic about the gauge coupling of the $\CN=4$ theory and consider the possible configurations of a state created by a $(p,q)$ string hanging between two pairs of branes.

In the first case we have  a $(p,q)$ string connecting the brane at the origin and that at $\tilde{z}_1$. This has a central charge 
\be
Z_{(p,q),1} \propto (p + q \tau) \tilde{z}_1 \ .
\ee 
Substituting from \eqref{eq:SU(3)rootmatching1}, get
\be
\begin{split}
Z_{(p,q),1} &\propto -\omega^2 (p + q \tau)(z_1 - \omega^2 z_2 ) \ , \\
            &\propto ( p + p \omega - q \omega^2 \tau  )(z_1 - \omega^2 z_2 ) \ .
\end{split}
\ee
This can be interpreted as some $(p',q')$ string suspended between the first brane (located at $z_1$) and the second image, located at $\omega^2 z_2$, of the second brane, iff $\tau=\omega^n$. More specifically, we get $p'=p-q$ and $q'=p$, if we use the value $\tau=\omega$ as suggested in   \cite{Aharony:2016kai}.

In the second case a $(p,q)$ string is suspended between the origin and $\tilde{z}_2$ and it
creates a state with central charge 
\be
\begin{split}
Z_{(p,q),2} &\propto (p + q \tau) \tilde{z}_2 \, \\
            &\propto \omega (p+q\tau) (z_1-\omega z_2) \ , \\
            &\propto (p \omega + q\omega \tau)(z_1-\omega z_2) \ .
\end{split} 
\ee 
Again, this looks like a $(p',q')$ string stretched between $z_1$ and the image brane at $\omega z_2$ iff $\tau=\omega^n$. For $\tau=\omega$, this gives $p'=-q$ and $q'=p-q$.

In the third case we have a $(p,q)$ string stretched between $\tilde{z}_1$ and $\tilde{z}_2$, it yields  
\be
\begin{split}
	Z_{(p,q),3} &\propto (p + q \tau)(\tilde{z}_1-\tilde{z}_2) \, \\
	            &\propto (p+q\tau) (z_1-z_2) \ . 
\end{split} 
\ee
This is equivalent to a $(p,q)$ string hanging between $z_1$ and $z_2$ when $\tau=\omega$.

Henceforth, we will choose the gauge coupling of the $\CN=4$ theory to be $\tau=\omega$. This is also the value of the complex structure of the F-theory torus in the $S_{3,1}$-fold. This implies that the LHS and RHS in \eqref{eq:SU(3)rootmatching1} become equal instead of being merely proportional i.e.
\be
\begin{split}
\tilde{z}_1 - \tilde{z}_2 & = z_1-z_2 \ , \\
              \tilde{z}_2 &= \omega z_1 + (1+\omega) z_2 = \omega (z_1-\omega z_2) \ , \\
              \tilde{z}_1 &= -\omega^2 z_1 + \omega z_2 = -\omega^2 (z_1 - \omega^2 z_2 ) \ .  
\end{split}
\label{eq:SU(3)rootmatching}
\ee

We now consider a state in the $S_{3,1}$-fold set-up, such that  its electromagnetic charge vector is  $(n_e^1,n_m^1;n_e^2,n_m^2)$. Its central charge then becomes \footnote{Here by $g_{YM}$ in $S_{k,l}$-fold, we mean the constant extracted from substituting the string coupling in \eqref{eq:StringCoupling}. }
\be
Z=g_{YM}\Big((n_e^1 + \tau n_m^1) z_1 + (n_e^2 + \tau n_m^2) z_2 \Big) 
\ee 
Let this correspond to a state with charges $(\tilde{n}_e^1,\tilde{n}_m^1; \tilde{n}_e^2,\tilde{n}_m^2)$ in the $\CN=4$ $SU(3)$ theory. In this description, its central charge is 
\be
Z=g_{YM}\Big((\tilde{n}_e^1 + \tau \tilde{n}_m^1) \tilde{z}_1 + (\tilde{n}_e^2 + \tau \tilde{n}_m^2) \tilde{z}_2 \Big) \ ,
\ee 
It is important to note that the coupling $\tau$ and hence $g_{YM}$ is same in both of our set-ups above. Requiring the equality of the central charges and using the relations in \eqref{eq:SU(3)rootmatching} (and their inverse) along with the fact that $\tau = \omega$, we find the following relations between the electromagnetic charges of a given state
\be
\begin{split}
	n_e^1 &= \tilde{n}_e^1 -\tilde{n}_m^1 -\tilde{n}_m^2 , \ \  n_m^1 = \tilde{n}_e^1 +\tilde{n}_e^2 -\tilde{n}_m^2\ , \\
	n_e^2 &= \tilde{n}_e^2 -\tilde{n}_m^1 -\tilde{n}_m^2 , \ \  n_m^2 = \tilde{n}_e^1 +\tilde{n}_e^2 -\tilde{n}_m^1 \ ,
\end{split}
\ee  
or equivalently
\be
\begin{split}
  \tilde{n}_e^1 &= \frac{1}{3}(n_e^1+n_m^1-2n_e^2 + n_m^2), \ \ \tilde{n}_m^1 = \frac{1}{3}(-n_e^1 + 2n_m^1-n_e^2 -n_m^2) \ ,\\
  \tilde{n}_e^2 &= \frac{1}{3}(-2 n_e^1+n_m^1+n_e^2 + n_m^2), \ \ \tilde{n}_m^2 = \frac{1}{3}(-n_e^1 - n_m^1-n_e^2 +2 n_m^2) \ .\\
\end{split}
\label{eq:chargesfrom3to4}
\ee

\section{3-pronged string in $S$-fold background}
\label{3-pronged}

In this section we refine our discussion on the enhacement of the rank 2 $S_{3,1}$-fold 
$\mathcal{N}=3$ theory to $\mathcal{N}=4$ by introducing a new character in the game, namely the 3-pronged string (a.k.a. 3-string junction or simply a 3-string).

As a warm up we first discuss the mass of an F1 string in the S-fold geometry (the case of a D1 is completely analogous). 
In the second part of the section we use the same formalism to compute 
the mass of the three pronged junction and we compare it with the central charge, obtained independently.

We then go on to study the walls of marginal stability and 
wall crossing in the S-fold background and find its analogue in the $\CN=4$ picture . 
We conclude this section by studying the behavior of 3-pronged junctions
in the flat geometry probed by three D3 branes, 
 and show how it fits with our understanding of $S$-folds.

\subsection{Warm-up: $M_{BPS}$ of an F1 string probing the rank 2 $S_{3,1}$ geometry}

Let us first consider a system of two D3 branes probing an $S_{3,1}$-fold. 
We require the branes to be coplanar, with their positions being $z_1'$ and $z_2'$ respectively. Their images are located at $\omega^n z_1'$ and $w^n z_2'$ respectively, with $n=1,2$ and $\omega^3=1$. In this setup $\tau = \omega = -\half + \frac{\sqrt{3}}{2} i $.  

\begin{figure}
	\centering
	\begin{subfigure}[t]{2.9in}
		\centering
		\includegraphics[height=1.0in]{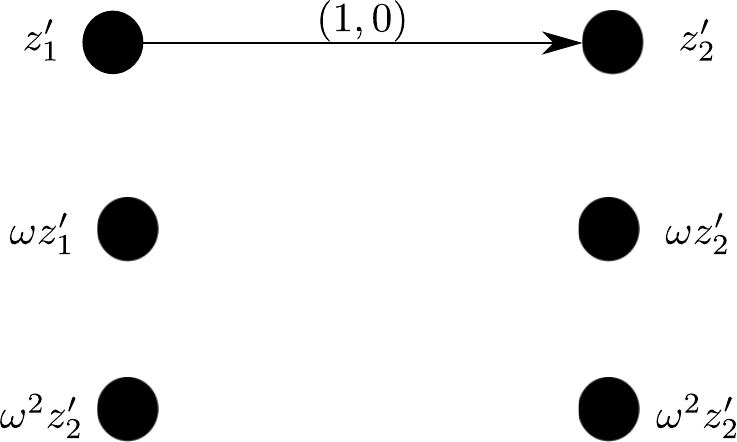}
		\caption{$(1,0)$ string between the two branes.}
		\label{fig:StringOnSfold1}
	\end{subfigure}
	\begin{subfigure}[t]{2.9in}
		\centering
		\includegraphics[height=1.0in]{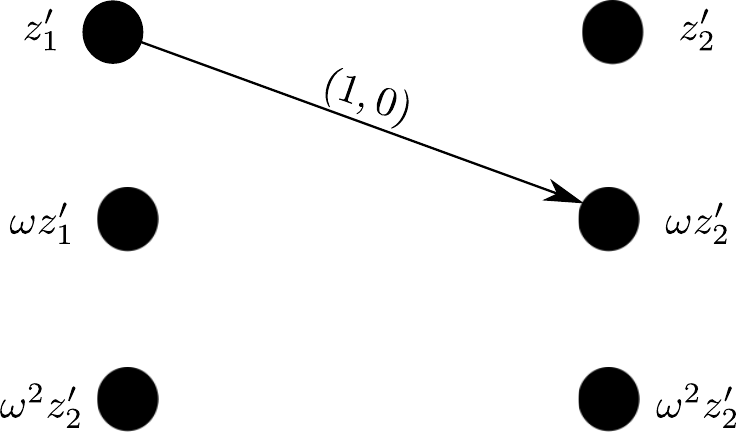}
		\caption{$(1,0)$ string between the first brane and an image of the second brane.}
		\label{fig:stringOnSfold2}
	\end{subfigure}
	\caption{(1,0) strings stretching between D3 branes in the S-fold geometry.}
	\label{fig:stringsOnSfold}
\end{figure}

Here we consider the mass of a (1,0)-string (an F1 string) suspended between the two branes (see figure \ref{fig:StringOnSfold1}).
The mass of the string is then given by 
\be
M^2_{F-string} =  \frac{g_s}{(2 \pi \alpha')^2} |z_1'-z_2'|^2 = 4 \pi g_s |z_1-z_2|^2 \ ,
\ee
where we used the relation $z'_a= 4 \pi^{\frac{3}{2}} \alpha' z_a $. This is expected to correspond to the state with electromagnetic quantum numbers $(n_e^1,n_e^2)=(1,-1)$ and $n_m^a=0$. The corresponding electric and magnetic charges are 
\be
\begin{split}
\vec{Q_E} &= g_{YM}( z_1 - z_2 ) \ , \\
\vec{Q_M}&=0 \ .
\end{split}
\ee
By using this result the mass of the corresponding BPS state is
\be
M^2_{BPS} = g_{YM}^2 |z_1-z_2|^2 \ .
\ee
This matches with the mass of the string after using the fact that $g_{YM}^2 = 4 \pi g_s$.

Similarly we can consider an F1 string suspended between the first brane and the first image of the second brane (figure \ref{fig:stringOnSfold2}). This implies that the end points of the string are at $z_1'$ and $\omega z_2'$ respectively and its mass if given by 
\be
M^2_{F-string} =  \frac{g_s}{(2 \pi \alpha')^2} |z_1'-\omega z_2'|^2 = 4 \pi g_s |z_1- \omega z_2|^2 \ ,
\ee
The end of the string on the first brane gives rise to the quantum numbers $(n_e^1=1, n_m^1=0)$. The other end of string ends on the image of the second brane. A $(p,q)$string ending on the $n$-th image will impart quantum numbers given by $ (n_e^a + n_m^a \tau)z_a =  -(p+q \tau) w^n z_a$ (without any sum over $a$). Thus, in the present case the other end of the string gives us $(n_e^2,n_m^2) = (0,-1)$. Thus we expect the corresponding state to have $(n_e^1 =1,n_e^2=0,n_m^1=0,n_m^2=-1)$ and 
\be
\begin{split}
\vec{Q_E} &= g_{YM} \big(z_1 +(-\half) (-z_2) \big)  =  g_{YM} \big(z_1 + \half z_2 \big) \ , \\
\vec{Q_M} & = -\frac{4 \pi}{g_{YM}} z_2 \ , 
\end{split}
\ee
here we have used the fact that $\frac{\theta}{2\pi} = -\half$. The mass of the corresponding BPS state is then given by 
\be
\begin{split}
M^2_{BPS} & = |Z|^2 = g_{YM}^2 |z_1 + \half z_2|^2 +  \big(\frac{4 \pi}{g_{YM}}\big)^2 |z_2|^2 +  8 \pi |\rm{Im} \big(z_1 + \half z_2 \big) z_2^*| \ .
\end{split}
\ee
Here we have used the fact that if we write every real 2-vector $(a,b)$ as the complex number $a+ i b$,  then $|\vec{v_1} \times \vec{v_2}| = |\rm{Im} ( \vec{v_1} \vec{v_2}^*)|$. The above expression can be further simplified by using $\frac{4 \pi}{g_{YM}^2} = \frac{\sqrt{3}}{2}$ to get
\be
M^2_{BPS} = |Z|^2 = g_{YM}^2 |z_1 - \omega z_2|^2 \ .
\ee
This matches with the mass of the string after using $g_{YM}^2 = 4 \pi g_s$.

Similarly, we can check that the mass of a D1 string hanging between the first brane and the first image of the second brane matches with the mass of a particle with quantum numbers $(n_e^1=0,n_e^2=1,n_m^1=1,n_m^2=1)$

\subsection{The case of the 3-pronged string}
\label{sec:3-strings}

\begin{figure}
	\centering
	\begin{subfigure}[t]{2.9in}
		\centering
		\includegraphics[height=2.9in]{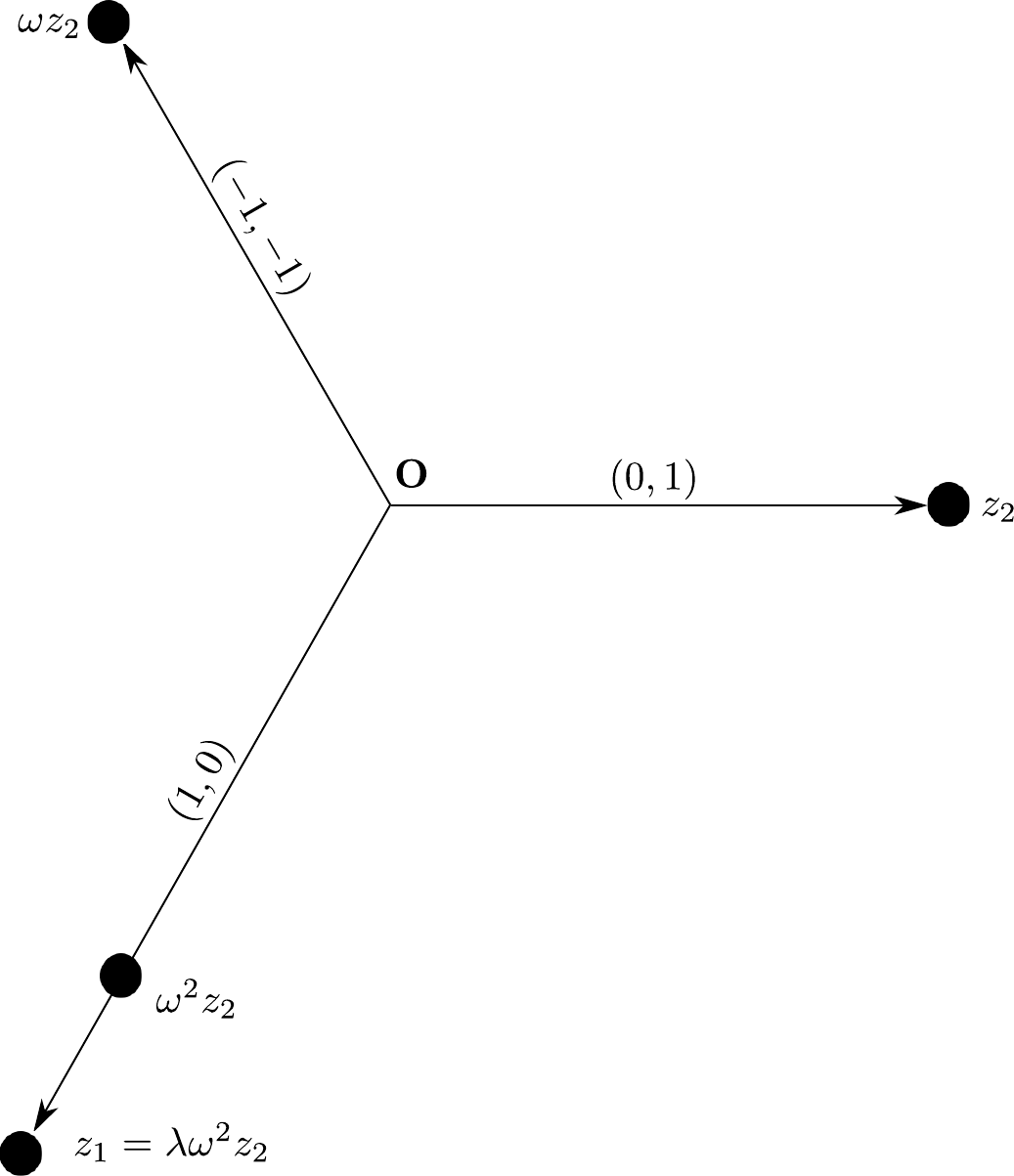}
		\caption{The vertex of the 3-string is at the origin.}
		\label{fig:3StringOnSfold1}
	\end{subfigure}
	\begin{subfigure}[t]{2.9in}
		\centering
		\includegraphics[height=2.9in]{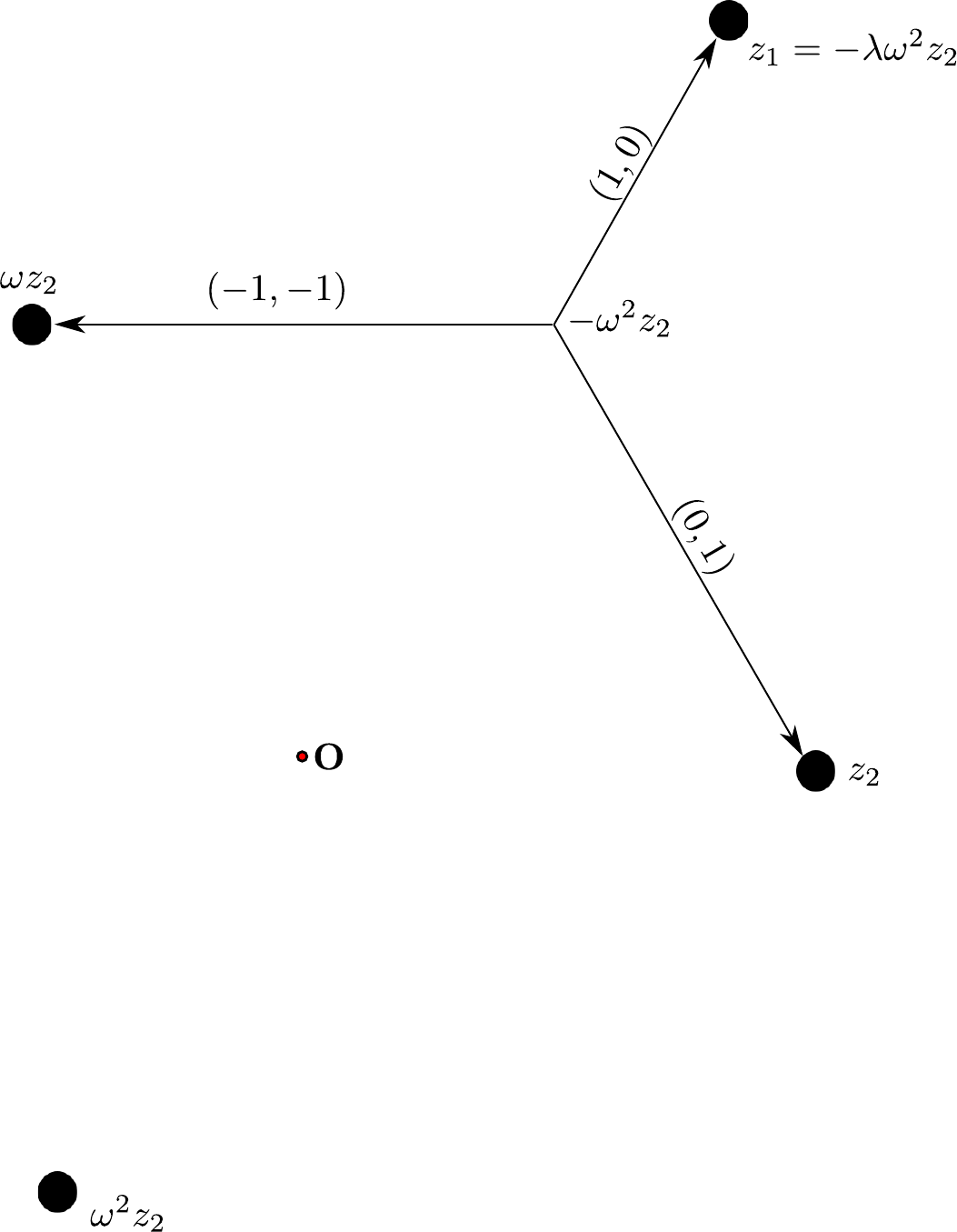}
		\caption{The vertex of the 3-string is at $-\omega^2 z_2$.}
		\label{fig:3stringOnSfold2}
	\end{subfigure}
		\caption{The two ways in which a 3-string with $(1,0),(0,1)$ and $(-1,-1)$ prongs can be made to end on the first brane, the second brane and the first image of the second brane. }
		\label{fig:3stringsOnSfold}
\end{figure}

Let us now consider a  3-string whose prongs are composed of $(1,0)$,  $(0,1)$ and $(-1,-1)$ strings respectively. Using \eqref{eq:stringtension} and $\tau=\omega$, we find that the tension of each prong is $\frac{\sqrt{g_s}}{2 \pi \alpha'}$. Cancellation of forces at the 3-string vertex then requires that the 3-string be oriented such that the angle between any pair of prongs is  $\frac{2 \pi}{3}$ radians. A possible way to suspend this string between the branes in the $S_{3,1}$-fold is given by ending the $(1,0)$-prong on the first brane (located at $z_1'$), ending the $(0,1)$-prong on the second brane (located at $z_2'$) and ending the $(1,1)$-prong on the first image of the second brane (located at $\omega z_2 '$). There are only two possible ways in which such a configuration can exist (see figure \ref{fig:3stringsOnSfold}). This is because the prongs of the string must enclose and angle of $\frac{2\pi}{3}$ rad. between them. This implies that If we consider the prongs ending on the second brane and its image then, the only way these two prongs can enclose an angle of  $\frac{2\pi}{3}$ rad. is when the vertex of the 3-string is either at the origin or at the point $-\omega^2 z_2$.

 When the 3-string vertex is located at the origin, the $(1,0)$ prong lies along the position vector of $\omega^2 z_2$. Therefore, it must be that $z_1' = \lambda \omega^2 z_2'$ with $\lambda\geq 0$ (figure \ref{fig:3StringOnSfold1}). The length of the prongs are $\lambda |z_2'|$, $|z_2'|$ and $|\omega z_2'|$ respectively. This implies that the mass of the string is 
\be
M_{3-string}^2 =  \frac{g_s}{(2 \pi \alpha')^2}(\lambda + 2)^2 |z_2'|^2 = 4 \pi g_s (\lambda + 2)^2 |z_2|^2 \ .
\label{eq:3-stringMass1}
\ee 
The quantum numbers of the corresponding field theory state can be calculated by considering the following: The $(1,0)$-prong ending on the first brane imparts $(n_e^1=1,n_m^1=0)$, the $(0,1)$-prong ending on the second prong imparts $(n_e^2=0,n_m^2=1)$ while, as explained earlier, the $(-1,-1)$-string ending on the first image of the second brane imparts $(n_e^2=1,n_m^2=0)$. Thus, we expect the net quantum numbers to be
$(n_e^1=1,n_e^2=1,n_m^1=0,n_m^2=1)$.  The corresponding charges are 
\be
\begin{split}
\vec{Q_E} &= g_{YM} \big( z_1+z_2 -\half z_2 \big) = g_{YM} \big( z_1+\half z_2 \big) \\
                 & =g_{YM} \big( \lambda \omega^2 +\half \big) z_2 \ , \\
\vec{Q_M} & = \frac{4 \pi}{g_{YM}} z_2  \ .                                
\end{split}
\label{eq:3stringcharges}
\ee
The mass and central charge of the BPS state is therefore given by
\be
\begin{split}
M^2_{BPS} = |Z|^2 &= g_{YM}^2 |z_2|^2 \Big( |\lambda \omega^2 +\half |^2 +  (\frac{4 \pi}{g_{YM}^2})^2 + 2 \frac{4 \pi}{g^2_{YM}} |\rm{Im} (\lambda \omega^2 +\half )| \Big)  \\
                             &=g_{YM}^2 |z_2|^2 \Big( \frac{1}{4}(1-\lambda)^2+ \frac{3}{4} \lambda^2 + \frac{3}{4} + \frac{3\lambda}{2} \Big) \\
                             &=g_{YM}^2 |z_2|^2 \Big(\lambda^2 + \lambda + 1 \Big) \ .
\end{split}
\label{eq:BPSmass1}
\ee
It is easy to check that there is no value of $\lambda > 0$ for which the masses  in \eqref{eq:3-stringMass1} and \eqref{eq:BPSmass1} agree with each other.  
From the map \eqref{eq:chargesfrom3to4}, between the charges in the S-fold and the $\CN=4$ theory, we will see that this should correspond to a magnetic monopole of the $\CN=4$ theory. Since monopoles are always $1/2$-BPS with respect to $\CN=4$ SUSY, they can not arise from non-BPS objects of the manifestly $\CN=3$ set-up. This is because the fact that it is non-BPS with respect to $\CN=3$ algebra implies that all the three supercharges of the $\CN=3$ algebra act non-trivially on this multiplet which in contradiction with the fact that only two of the four supercharges of the $\CN=4$ algebra can act non-trivially on a $1/2$-BPS multiplet. We therefore \emph{conjecture} that such string configurations can not exist in the S-fold. 

Similarly, when the 3-string junction is located at $-\omega^2 z_2$, as shown in figure \ref{fig:3stringOnSfold2}, the $(1,0)$ prong lies along the position vector of $-\omega^2 z_2$ and hence the first brane must be located at $z_1' = -\lambda \omega^2 z_2' $ with $\lambda\geq 1$. The length of the prongs will then be $(\lambda-1)|z_2'|$, $|z_2'|$ and $|\omega z_2'|$. The mass of the string in this configuration is 
\be
M_{3-string}^2 =  \frac{g_s}{(2 \pi \alpha')^2}(\lambda + 1)^2 |z_2'|^2 = 4 \pi g_s (\lambda + 1)^2 |z_2|^2 \ .
\label{eq:3-stringMass2}
\ee 
Changing the position of the 3-string junction does not affect the electromagnetic quantum numbers that the end of the strings create on the brane. Thus, we can calculate the mass of the corresponding CFT state by taking $\lambda \rightarrow -\lambda$ in \eqref{eq:3stringcharges}. This gives the mass of the BPS state to be 
\be
\begin{split}
	M^2_{BPS} =|Z|^2 &= g_{YM}^2 |z_2|^2 \Big( |-\lambda \omega^2 +\half |^2 +  (\frac{4 \pi}{g_{YM}^2})^2 + 2 \frac{4 \pi}{g^2_{YM}} |\rm{Im} (-\lambda \omega^2 +\half )| \Big)  \\
	&=g_{YM}^2 |z_2|^2 \Big( \frac{1}{4}(1+\lambda)^2+ \frac{3}{4} \lambda^2 + \frac{3}{4} + \frac{3\lambda}{2} \Big) \\
	&=g_{YM}^2 |z_2|^2 (\lambda+ 1)^2 \ .
\end{split}
\label{eq:BPSmass2}
\ee
This matches with \eqref{eq:3-stringMass2} for all values of $\lambda$ and represents a $1/3$-BPS state with respect to the manifest $\CN=3$ SUSY algebra. Using \eqref{eq:chargesfrom3to4}, we can now map the charge vector, $(n_e^1=1,n_m^1=0;n_e^2=1,n_m^2=1)$, of this state from the S-fold geometry to that in the $\CN=4$ $SU(3)$ gauge theory where it becomes $(\tilde{n}_e^1=0,\tilde{n}_m^1=-1;\tilde{n}_e^2=0, \tilde{n}_m^2=0)$. We therefore infer that in the $SU(3)$ gauge theory described by 3 D3 branes on a flat background, the 3-string of figure \ref{fig:3stringOnSfold2} corresponds to a $(0,-1)$ string stretched between the brane at the origin and the brane at $\tilde{z}_1$ i.e. it represents a 1/2-BPS magnetic monopole of the $SU(3)$ theory. For the number of degrees of freedom to match in the two descriptions, it must therefore be that the hidden supercharge of the S-fold set-up acts trivially on the 3-string of the figure \ref{fig:3stringOnSfold2}.
\subsection{Wall crossing}
\begin{figure}
	\centering
	\begin{subfigure}[t]{2.9in}
		\centering
		\includegraphics[height=1.5in]{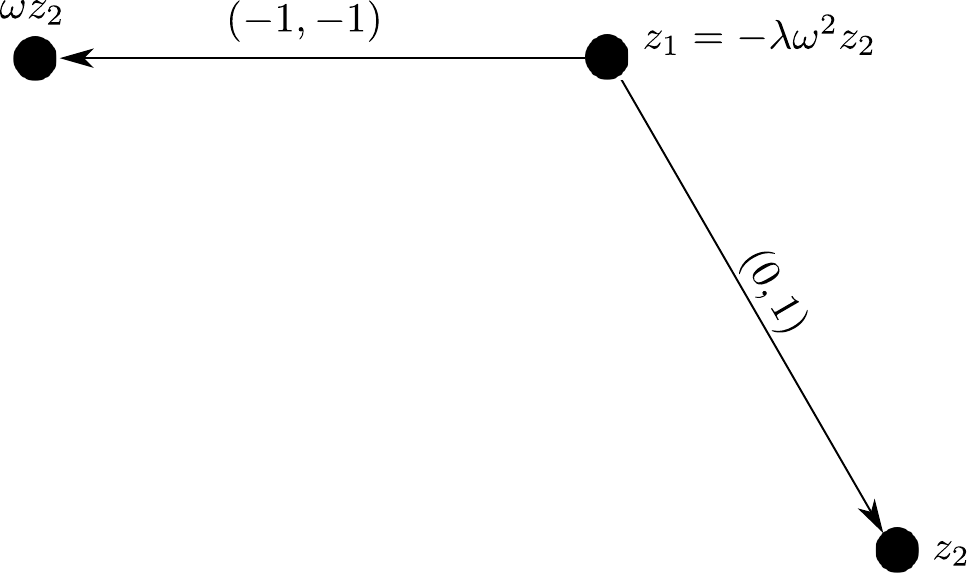}
		\caption{Wall crossing in S-fold background.}
		\label{fig:SfoldWallCrossing}
	\end{subfigure}
	\begin{subfigure}[t]{2.9in}
		\centering
		\includegraphics[height=1.5in]{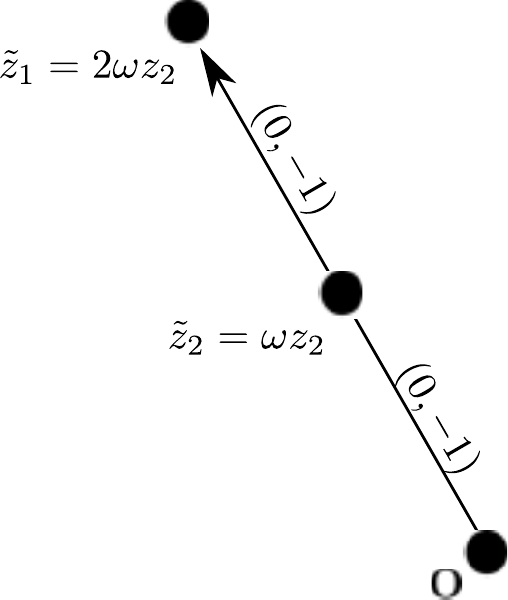}
		\caption{The analogue of figure \ref{fig:SfoldWallCrossing} in flat background}
		\label{fig:FlatWallCrossing}
	\end{subfigure}
		\caption{Wall of marginal stability in the S-fold and flat backgrounds. }
		\label{fig:WallCrossing}
\end{figure}

We can also consider walls of marginal stability in the above scenario.  For the 3-string of figure \ref{fig:3stringOnSfold2}, this corresponds to setting $\lambda=1$ such that $z_1=-\omega^2 z_2$. Doing this reduces the $(1,0)$ prong to zero length, the 3-string configuration now consists of a $(-1,-1)$ string stretched from the brane at $z_1=-\omega^2 z_2$ to the brane at $\omega z_2$ and a $(0,1)$ string stretched from the brane at $z_1=-\omega^2 z_2$ to the brane at $z_2$ as shown in figure \ref{fig:SfoldWallCrossing}. These two string can now move independently of each other, thereby giving rise to wall crossing. Using \eqref{eq:SU(3)rootmatching}, to map to the manifestly $\CN=4$ set-up, we find that the three branes in flat background are now located at the origin, $\tilde{z}_1=2 \omega z_2$ and $\tilde{z}_2 = \omega z_2$, respectively. This implies that the brane at $\tilde{z}_2$ is co-linear with and lies mid-way between the brane at the origin and that at $\tilde{z}_1$ as shown in figure \ref{fig:FlatWallCrossing}.
 The $(0,-1)$ stretched between the latter two branes can now break on the brane at $\tilde{z}_2$ hence triggering wall-crossing in the flat geometry. This can be further tested by noting that the state corresponding to a $(-1,-1,)$ string stretched between $z_1$ and $\omega z_2$ has a charge vector given by $(n_e^1=1,n_m^1=1; n_e^2=1,n_m^2=0)$. Using \eqref{eq:chargesfrom3to4}, we find that  this corresponds to a state with charge vector $(\tilde{n}_e^1=0,\tilde{n}_m^1=0;\tilde{n}_e^2=0, \tilde{n}_m^2=-1)$ in the flat geometry i.e. a $(0,-1)$ string going from the brane at the origin to the brane at $\tilde{z}_2$. Similarly, in the S-fold, the $(0,1)$ string from $z_1$ to $z_2$ gives a state with charges $(n_e^1=0,n_m^1=-1; n_e^2=0,n_m^2=1)$. The corresponding state in $\CN=4$ $SU(3)$ SYM has charges  $(\tilde{n}_e^1=0,\tilde{n}_m^1=-1;\tilde{n}_e^2=0, \tilde{n}_m^2=1)$ which is same as that due to a $(0,-1)$ string going from $\tilde{z}_2$ to $\tilde{z}_1$. 
 
We can also consider the scenario when $\lambda$ in Figure \ref{fig:3stringOnSfold2} is tuned to be such that $0 < \lambda < 1$. As is evident from the discussion so far, the 3-string does not exist for these values of $\lambda$. In stead it undergoes wall crossing. We therefore end up with a  $(-1,-1,)$ string stretched between the branes at $z_1$ and $\omega z_2$ and a $(0,1)$ string from the brane at $z_1$ to that at $z_2$. On the manifestly $\CN=4$ side, this will be described by a $(0,-1)$ string going from the brane at the origin to the brane at $\tilde{z}_2$ and another $(0,-1)$ string going from $\tilde{z}_2$ to $\tilde{z}_1$

\subsection{3-pronged strings in $\CN=4$ $SU(3)$ theory.}
It is also interesting to consider the $1/4$-th BPS states of $\CN=4$ $SU(3)$ theory generated by 3-strings as was done in \cite{Bergman:1997yw}. We therefore consider 3-string with its prongs having charges: $(1,0), \ (0,1), \ (-1,-1)$. Let us consider the following configuration: the prong with charge $(0,1)$ ends on the brane at $\tilde{z}_1$, the prong with charge $(1,0)$ ends on the brane at $\tilde{z}_2$ and the prong with charge $(-1,-1)$ ends on the brane at the origin. This corresponds to a state with electromagnetic charge vector $(\tilde{n}_e^1=0,\tilde{n}_m^1=1;\tilde{n}_e^2=1, \tilde{n}_m^2=0)$. Using 
\eqref{eq:chargesfrom3to4}, we see that the corresponding charges in the S-fold geometry will be given by $(n_e^1=-1,n_m^1=1;n_e^2=0,n_m^2=0)$. These charges correspond to a $(-1,0)$ string stretching from the brane at $z_1$ to its image in the S-fold at $\omega z_1$.

At this point, we wish to remind ourselves of our discussion in section \ref{sec:Countingdof}, where we had stated that for $l=1$ in S-folds, there can be no $\CN=3$ vector multiplets corresponding to $\pm(1,0)$ strings stretched between a brane and its images in the S-fold. This is consistent with the above correspondence between the 3-string of $\CN=4$ $SU(3)$ theory  and the $(-1,0)$ string stretched from the brane at $z_1$  to its image at $\omega z_1$. This is because the correspondence tells us that the multiplet associated to the $(-1,0)$ string contains excitations with spin $>1$ and therefore it is larger than the $\CN=3$ vector multiplet.

\section{Further directions and open questions}

\label{sec:further}

In this paper we studied the enhancement of supersymmetry in rank 2 
$S_{3,1}$-fold geometry, from $\mathcal{N}=3$ to $\mathcal{N}=4$.
We developed a dictionary between $(p,q)$  strings in the S-fold geometry and the 
corresponding states associated to the flat $\mathcal{N}=4$ geometry.
This allowed us to compute masses and charges also for more complicated 
3-strings and compare the walls of marginal stability between the two descriptions.
One of the main difficulties arising in the analysis is related to the structure of the 
central charges in the $\mathcal{N}=3$ and in the $\mathcal{N}=4$ algebras.
Let us elaborate more on this point.

In theories with $\CN > 1$, we can extend the superalgebra by introducing central charges $Z^{ab}$. The SUSY algebra becomes 
\be
\begin{split}
\{ Q_{\a}^a,\bar{Q}_{\dot{\a} b}\} & = 2 \delta^a_b\sigma_{\a\dot{\a}}^\mu P_\mu  \ , \\
\{Q_{\a}^a,Q_{\b}^b\} &=2 \epsilon_{\a \b}Z^{ab} \ , \\
\{\bar{Q}_{\dot{\a} a},\bar{Q}_{\dot{\b} b}\} &=2 \epsilon_{\dot{\a} \dot{\b}}Z^{\dagger}_{\phantom{a}ab} \ .
\end{split}
\ee
 Here $\alpha, \ \dot{\a}$ are space-time spinor indices while $a,b$ run from 1 to $\CN$. From this we see that $Z^{ab}$ transforms in the two index antisymmetric representation of the $SU(\CN)_R$ symmetry of the algebra and when $\CN < 4$, it also also carries an R-charge 2 with respect to the $U(1)_R$ symmetry. We can now use the $SU(\CN)_R$ R-transformations to skew-diagonalize the central charge matrix of a given SUSY multiplet \footnote{We remind the reader that since central charge commutes with the supercharges, therefore all the components of a SUSY multiplet have the same central charge.}.  When, $\CN<4$, we can further use the $U(1)_R$ transformations to cancel the phase of one of the charges in the central charge matrix. This implies that for $\CN=2,3$, there is single real parameter that appears in the BPS condition while when $\CN=4$, there are two different central charges formed from 3 real parameters \footnote{The three real parameters controlling the central charges of $\CN=4$  SUSY are given by the magnitudes, $|\vec{Q_E}|$ and $|\vec{Q_M}|$, of the electric and the magnetic charge vectors and the angle $\varphi$ between them. }. In the rest frame of a particle of mass $M$, the supercharges can be then expressed in terms of linear combinations of operators $a_\a \ , \ b_\a \ , \ c_\a \ , \ d_\a$ and their conjugates, such that when $\CN=2$, the algebra becomes
 \be
 \begin{split}
 \{ a_\a, a_\b \} &=  \{ b_\a, b_\b \} = \{ a_\a, b_\b \} = 0 \ , \\
 \{ a_\a, (a_\b)^{\dagger} \} &= 2 \delta_{\a \b} (M+Z) \ , \\
  \{ b_\a, (b_\b)^{\dagger} \} &= 2 \delta_{\a \b} (M-Z) \ ,
 \end{split}
 \label{eq:N=2central}
 \ee
 while $c_\a$ and $d_\a$ will be absent. When $\CN=3$, the algebra of \eqref{eq:N=2central} will have to be augmented by the operator $c_\a$ obeying anti-commutation relations given by 
 \be
 \begin{split}
 \{ c_\a, c_\b \} &=  \{ c_\a, a_\b \} = \{ c_\a, b_\b \} = 0 \ , \\
 \{ c_\a, (c_\b)^{\dagger} \} &= 2 \delta_{\a \b} M \ .
 \end{split}
 \label{eq:N=2central}
 \ee
On the other hand when $\CN=4$, we will also have operators $c_\a$ and $d_\a$ in addition to those in \eqref{eq:N=2central}. Their algebra will then be given by 
 \be
 \begin{split}
  \{ c_\a, a_\b \} &= \{ c_\a, b_\b \} =   \{ d_\a, a_\b \} = \{ d_\a, b_\b \}=0 \ , \\
 \{ c_\a, c_\b \} &=  \{ d_\a, d_\b \} = \{ c_\a, d_\b \} = 0 \ , \\
 \{ a_\a, (a_\b)^{\dagger} \} &= 2 \delta_{\a \b} (M+\widetilde{Z}) \ , \\
  \{ b_\a, (b_\b)^{\dagger} \} &= 2 \delta_{\a \b} (M-\widetilde{Z}) \ ,
 \end{split}
 \label{eq:N=4central}
 \ee
where $Z$ and $\widetilde{Z}$ are the central charges obtained after skew-diagonalizing $Z^{ab}$. Since  the charges $Z$ and $\widetilde{Z}$ in $\CN=4$ systems, will generically have different magnitudes, the BPS condition can only be satisfied for the larger of the two of them. The $\CN=4$ BPS states of this kind will then preserve a single set of SUSY generators and their multiplet will be generated by the action of the other 3 supersymmetries that continue to act non-trivially when the BPS condition is saturated. Such BPS states are called 1/4-BPS and they have a total of $2^6$ d.o.f  (bosonic + fermionic) in their  multiplets. However, when a state is such that both the central charges are equal, then the BPS condition implies that the state remains invariant under two of the four supercharges, giving us a $1/2$-BPS multiplet with a total of $2^4$ on-shell d.o.f . Note that all the multiplets of $\CN=4$ SUSY are CPT self-conjugate. 

On the other hand, in $\CN=2,3$ theories, there is single BPS condition and therefore, the action of only a single set of SUSY generators can become trivial. This implies that the BPS states can be at most 1/2-BPS and 1/3-BPS respectively. Generically, the total number of degrees of freedom in these multiplets will be $2^3$ when $\CN=2$ (after including the CPT conjugate states) and $2^4$ when $\CN=3$ (this is a CPT self-conjugate multiplet).  

We now consider those $\CN=3$ set-ups that were argued to have an accidental enhancement to $\CN=4$ SUSY. In this case the theory possesses an extra hidden supercharge as well as an extra hidden central charge. The ratio of mass to the hidden central charge then tells us if the BPS states will become 1/2-BPS or 1/4-th BPS states in the $\CN=4$ theory. 
The counting of total number of d.o.f in the 1/3-BPS states of $\CN=3$ matches with that of the 1/2-BPS states of $\CN=4$ theories. From this, it is clear that upon SUSY enhancement, a manifestly 1/3-BPS states of the $\CN=3$ theory will simply uplift to the $\CN=4$ 1/2-BPS states iff they are invariant under the extra hidden supercharges. This also implies that the BPS bound with respect to the hidden central charge will get saturated. On the other hand the $\CN=4$ 1/4-BPS states have more d.o.f than the $\CN=3$ 1/3-BPS states. This implies the 1/3-BPS states can not uplift directly to 1/4-BPS states of $\CN=4$. The hidden supercharge must therefore act non-trivially on such a BPS multiplet and pair it up with other $\CN=3$ multiplets to form an $\CN=4$ 1/4-BPS multiplet. In this case, the BPS bound with respect to the hidden central charge can not be saturated. It can also happen that a manifestly non-BPS multiplet of $\CN=3$ ends up becoming a 1/4-BPS multiplet iff the hidden supercharge acts trivially on it. In this case the BPS bound with respect to the manifest $\CN=3$ central charge is not saturated but that with respect to the hidden central charge does get saturated.  

As was already argued in section \ref{sec:3-strings}, a non-BPS multiplet of $\CN=3$ theory can never become a $1/2$-BPS multiplet upon SUSY enhancement to an $\CN=4$ theory. This is simply because the non-BPS nature of the multiplet implies that its components get transformed into each other under all the three supersymmetries in $\CN=3$ algebra while requiring it to be $1/2$-BPS with respect to enclosing $\CN=4$ algebra would imply that it is invariant under two of the four supersymmetries. Clearly, both these requirements can not be satisfied at the same time.  

We conclude the analysis by mentioning  another interesting problem that we did not discuss here. It regards the 
extension of our analysis to the other rank 2 cases leading to the non-perturbative enhancement of $\mathcal{N}=3$ SUSY to $\mathcal{N}=4$. Such cases correspond to the 
$S_{4,1}$-fold and to the $S_{6,1}$-fold.
As mentioned above they enhance to $SO(5)$ and $G_2$ $\mathcal{N}=4$ SYM respectively.
The first case has a perturbative realization in terms of D3 branes and an $\widetilde{O3}^{-}$ plane and it may be studied along the lines of our analysis.
On the other hand, the second case requires some more care, because a perturbative realization of the $G_2$ theory is absent in string theory. 
We hope to come back to these issues in the next future.

\section*{Acknoledgement}

We thank Sangmin Lee for initial collaboration on this project and for insightful comments on the draft. We are also grateful to Dario Rosa,  Jaewon Song, Kimyeong Lee and Seok Kim for useful discussions. 
The work of P.~A.~is supported by Samsung Science and Technology Foundation under Project Number SSTF-BA1402-08.
P.~A.~is grateful to the IPMU  for hospitality during the last stages of this work.
A.~A.~is grateful to the university of Milano-Bicocca for hospitality during the last stages of this work.
\bibliographystyle{ytphys}
\bibliography{refs}

\providecommand{\href}[2]{#2}\begingroup\raggedright\begin{thebibliography}{10}

\bibitem{Aharony:2015oyb}
O.~Aharony and M.~Evtikhiev, ``{On four dimensional N=3 superconformal
  theories},''
\href{http://arxiv.org/abs/1512.03524}{{\ttfamily arXiv:1512.03524 [hep-th]}}.

\bibitem{Garcia-Etxebarria:2015wns}
I.~Garcia-Etxebarria and D.~Regalado, ``{$ \mathcal{N}=3 $ four dimensional
  field theories},'' \href{http://dx.doi.org/10.1007/JHEP03(2016)083}{{\em
  JHEP} {\bfseries 03} (2016) 083},
\href{http://arxiv.org/abs/1512.06434}{{\ttfamily arXiv:1512.06434 [hep-th]}}.

\bibitem{Ferrara:1998zt}
S.~Ferrara, M.~Porrati, and A.~Zaffaroni, ``{N=6 supergravity on AdS(5) and the
  SU(2,2/3) superconformal correspondence},''
  \href{http://dx.doi.org/10.1023/A:1007592711262}{{\em Lett. Math. Phys.}
  {\bfseries 47} (1999) 255--263},
\href{http://arxiv.org/abs/hep-th/9810063}{{\ttfamily arXiv:hep-th/9810063
  [hep-th]}}.

\bibitem{Beck:2016lwk}
S.~W. Beck, J.~B. Gutowski, and G.~Papadopoulos, ``{AdS$_{5}$ backgrounds with
  24 supersymmetries},'' \href{http://dx.doi.org/10.1007/JHEP06(2016)126}{{\em
  JHEP} {\bfseries 06} (2016) 126},
\href{http://arxiv.org/abs/1601.06645}{{\ttfamily arXiv:1601.06645 [hep-th]}}.

\bibitem{Cordova:2016xhm}
C.~Cordova, T.~T. Dumitrescu, and K.~Intriligator, ``{Deformations of
  Superconformal Theories},''
\href{http://arxiv.org/abs/1602.01217}{{\ttfamily arXiv:1602.01217 [hep-th]}}.

\bibitem{Nishinaka:2016hbw}
T.~Nishinaka and Y.~Tachikawa, ``{On 4d rank-one N=3 superconformal field
  theories},''
\href{http://arxiv.org/abs/1602.01503}{{\ttfamily arXiv:1602.01503 [hep-th]}}.

\bibitem{Argyres:2016xua}
P.~C. Argyres, M.~Lotito, Y.~Lu, and M.~Martone, ``{Expanding the landscape of
  $ \mathcal{N} $ = 2 rank 1 SCFTs},''
  \href{http://dx.doi.org/10.1007/JHEP05(2016)088}{{\em JHEP} {\bfseries 05}
  (2016) 088},
\href{http://arxiv.org/abs/1602.02764}{{\ttfamily arXiv:1602.02764 [hep-th]}}.

\bibitem{Aharony:2016kai}
O.~Aharony and Y.~Tachikawa, ``{S-folds and 4d N=3 superconformal field
  theories},''
\href{http://arxiv.org/abs/1602.08638}{{\ttfamily arXiv:1602.08638 [hep-th]}}.

\bibitem{Morrison:2016nrt}
D.~R. Morrison and C.~Vafa, ``{F-Theory and N=1 SCFTs in Four Dimensions},''
\href{http://arxiv.org/abs/1604.03560}{{\ttfamily arXiv:1604.03560 [hep-th]}}.

\bibitem{Imamura:2016udl}
Y.~Imamura, H.~Kato, and D.~Yokoyama, ``{Supersymmetry Enhancement and
  Junctions in S-folds},''
\href{http://arxiv.org/abs/1606.07186}{{\ttfamily arXiv:1606.07186 [hep-th]}}.

\bibitem{Hull:2004in}
C.~M. Hull, ``{A Geometry for non-geometric string backgrounds},''
  \href{http://dx.doi.org/10.1088/1126-6708/2005/10/065}{{\em JHEP} {\bfseries
  10} (2005) 065},
\href{http://arxiv.org/abs/hep-th/0406102}{{\ttfamily arXiv:hep-th/0406102
  [hep-th]}}.

\bibitem{Dabholkar:2003xi}
A.~Dabholkar, ``{String compactifications: Old and new},'' in {\em {On recent
  developments in theoretical and experimental general relativity, gravitation,
  and relativistic field theories. Proceedings, 10th Marcel Grossmann Meeting,
  MG10, Rio de Janeiro, Brazil, July 20-26, 2003. Pt. A-C}}, pp.~148--164.
\newblock
2003.
\newblock

\bibitem{Nilse:2006jv}
L.~Nilse, ``{Classification of 1D and 2D orbifolds},''
\href{http://arxiv.org/abs/hep-ph/0601015}{{\ttfamily arXiv:hep-ph/0601015
  [hep-ph]}}.

\bibitem{Witten:1998xy}
E.~Witten, ``{Baryons and branes in anti-de Sitter space},''
  \href{http://dx.doi.org/10.1088/1126-6708/1998/07/006}{{\em JHEP} {\bfseries
  07} (1998) 006},
\href{http://arxiv.org/abs/hep-th/9805112}{{\ttfamily arXiv:hep-th/9805112
  [hep-th]}}.

\bibitem{Shephard01011952}
G.~C. Shephard, ``Regular complex polytopes,''
  \href{http://plms.oxfordjournals.org/content/s3-2/1/82.short}{{\em
  Proceedings of the London Mathematical Society} {\bfseries s3-2} no.~1,
  (1952) 82--97},
  \href{http://arxiv.org/abs/http://plms.oxfordjournals.org/content/s3-2/1/82.full.pdf+html}{{\ttfamily
  http://plms.oxfordjournals.org/content/s3-2/1/82.full.pdf+html}}.

\bibitem{shephard1953unitary}
G.~Shephard, ``Unitary groups generated by reflections,'' {\em CANADIAN JOURNAL
  OF MATHEMATICS-JOURNAL CANADIEN DE MATHEMATIQUES} {\bfseries 5} no.~3, (1953)
  364--383.

\bibitem{shephard1954finite}
G.~C. Shephard and J.~A. Todd, ``Finite unitary reflection groups,'' {\em
  Canad. J. Math} {\bfseries 6} no.~2, (1954) 274--301.

\bibitem{Bergman:1997yw}
O.~Bergman, ``{Three pronged strings and 1/4 BPS states in N=4 superYang-Mills
  theory},'' \href{http://dx.doi.org/10.1016/S0550-3213(98)00345-9}{{\em Nucl.
  Phys.} {\bfseries B525} (1998) 104--116},
\href{http://arxiv.org/abs/hep-th/9712211}{{\ttfamily arXiv:hep-th/9712211
  [hep-th]}}.

\end{thebibliography}\endgroup

\end{document}